Short Paper

# Assessment of E-Learning Readiness of Faculty Members and Students in the Government and Private Higher Education Institutions in the Philippines

Harold R. Lucero
School of Information Technology
Colegio De Sta. Teresa De Avila, Inc.
harold_ramirez_lucero@yahoo.com

Jayson M. Victoriano
Bulacan State University
jayson.victoriano@bulsu.edu.ph
(corresponding author)

Jennifer T. Carpio
School of Graduate Studies
UE Caloocan
jhennieffer@yahoo.com

Paquito G. Fernando Jr.
College of Computer Studies
John Paul College
kitzfernando@gmail.com

**Abstract**

*Purpose* – This study seeks to determine the level of readiness of selected private and government-managed colleges and universities in the Philippines. The study also aims to determine if there's a significant difference in the level of readiness between the private and government HEI's.

*Method* – In line with this, a descriptive-comparative method was employed. A questionnaire from the works of Aklaslan and Merca was utilized to determine the faculty members' and students' status of their e-learning readiness, acceptance, training, technological infrastructure, and tool awareness on the implementation of an e-learning program. Weighted mean was used to determine the level of readiness while a t-test was employed to determine if there is a significant difference between results.

*Results* – The finding shows that the assertions towards the readiness of implementing e-learning on both private and government-funded institutions show no significant difference and are generally accepted.

*Conclusion and Recommendation* – It can be concluded that higher education institutions are ready to adopt and implement e-learning towards continuous improvement in the quality of education in the Philippines. Thus, adoption and implementation of e-learning is recommended.

*Keywords* – E-learning, assertion, implementation, SUC, HEI

## INTRODUCTION

The Internet has become one of the most important tools in the 21st-century landscape of education. The use of the internet in all sectors has become a necessity and provides access to borderless resources. It is very impossible to remove technology from the equation today, even in the most remote places, because technology is making its way through every place in the world (Gomez, 2016). The continuous increase of emerging technologies also allows the creation of new paradigms for modern education, like an E-Leaning. E-learning is learning conducted via electronic means, typically on the internet. Through E-Learning, educators can provide information and instruction to students as a complement to face to face meetings. E-Learning can also be a solution for the cost efficiency and bring business competitiveness for educational institutions, including higher education (Laksitowening, Wibowo, & Hidayati, 2016).

In the implementation of E-learning in higher education institutions should not only focus on providing the required technology and making learning content available. The failure of E-Learning is often caused not by technological matters but the inability of educators and universities in providing the planning of the learning process and less attention to non-technical domains (Frimpon, 2012).

In implementing E-Learning, the following factors need to be considered: the readiness of the staff and administration, economic readiness, environmental readiness, technological readiness, and the readiness of the culture (Psycharis, 2005). There are many attempts in implementing e-learning and there are factors why initiative fails in both private and public sectors are: poor alignment to needs, communication, and lack of implementation skills, poor implementation process, management commitment, scalability, support, and technology (Sharma, Stone, & Ekinci 2009)

This study aims to assess the e-learning readiness of selected private and public higher education institutions and to determine if there's a significant difference in the level of readiness between the two, by using the criteria stated by previous studies. A survey instrument was utilized for the faculty members and students on the status of their e-learning readiness, acceptance, training, technological infrastructure, and tool awareness for implementing the e-learning program.

## RELATED WORKS

E-learning provides a platform that provides anywhere, Hassle-free access for up-gradation of knowledge and skills. It provides a platform wherein the individual gets a customized package related to key thematic areas, through a self-guided development. That means it is an association of technology, pedagogy, and accreditation to come up with the attractive concept of learning called e-Learning. It gives technology-enabled learning with the use of tolerable policy and various resources, such as text content, visual content, and audio content (Gowda & Suma, 2017).

Public and private Institutions have proposed a comprehensive and complete introduction to cloud computing. E-Learning based on cloud computing in higher education institutions as stressed in this paper could be utilized adequately, since it is a very cost-effective technology and also due to its manifold benefits. Also, cloud computing has led to the enhancement of the productivity and experience of I.T staff members in different institutions. According to (Alajmi et al. 2017), online education can provide a basic attendance check for day-to-day contact to be a gateway to effortless access to different applications. Consumerization, a trend of using and discovering technology for personal tasks and applications should be fully utilized in higher education institutions which can improve students' technology-enabled education and diminish conventional physical attendance for lecture meetings.

The objectives of the study were to analyze the relationship of university students' intention to use e-learning with selected constructs, such as their attitude, perceived usefulness, perceived ease of use, self-efficacy of e-learning, subjective norm and system accessibility, and to develop a general linear structural model of e-learning acceptance of

400

university students that would provide a School manager or an educator with implications for better-implementing e-learning (Park,2009).

The E-Learning approach on the market is limited to technical gadgets and organizational aspects of teaching. As a result, the learner has become de-individualized and demoted to a noncritical homogenous user. One way out of this drawback is the creation of an individual e-Learning management system, with a flexible multidimensional data model and the production of individual content are the solution (Tavangarian, 2004).

## METHODOLOGY

### Research Design

This study applied the descriptive research and quantitative approach to defining e-learning readiness towards the given factors, such as awareness on e-learning, acceptance, available technologies, familiarity in different web tools, and institution's facilities to support the implementation of such.

### Sample of the Population

The population of the respondents in this study is consist of ten (10) faculty members and forty (40) students from seven (7) different higher education institution. Four (4) of these institutions are managed by the government and the remaining three (3) are all privately owned colleges. A proportional random sampling method was used to obtain a population of the respondents.

### Instrument

The survey instrument used in this study was adopted from the combination of the work of Akaslan et al. (2012) and Mercado (2008) that provide factors used as an e-learning readiness assessment tool. The study used two sets of "E-Learning Readiness Survey": Faculty members, and students.

Viewpoint on technological infrastructure was taken on the study of Akaslan et al.'s. (2012), and a 5-point Likert-scale (ranging from "strongly disagree" to "strongly agree") was applied. It was coded 1,2,3,4, and 5, 5 being the highest and 1 is the lowest. Awareness and skills in web technologies were based on the combination of tools used by both Akaslan et al. (2012) and Mercado (2008). A dichotomous question was also employed to evaluate this area;

### Data Analysis

Weighted mean was used to determine the level of readiness while a t-test was used to determine if there is a significant difference between results. The Statistical Package for the Social Sciences (SPSS) version 20 was used to manipulate and analyze data to determine if there are significant differences between the level of readiness in implementing E-learning between private and government-funded institutions.

# RESULTS

Table 1 shows the e-learning readiness of faculty members and students in state universities and private universities in the Philippines. The table reveals that the students from both public and private universities demonstrated a high level of e-learning readiness as reflected in the computed mean of 4.24 from state universities and colleges (SUC) and 3.96 from private universities with a combined mean of 4.03 described as agreeing. Similarly, faculty from both the public and private universities registered a high level of e-learning readiness, with a weighted of 4.71 (SUC) and 4.09 (private) with a combined mean of 4.40 described as agreeing. The overall mean is 4.26 described as agree implies that faculty members and students from both private and public universities possess confidence in using e-infrastructure, such as the Internet, software, digital tools in teaching and learning. Moreover, they exhibited the necessary computer skills needed in the implementation of e-learning.

Table 1. E-Learning Readiness of Faculty Members and Students in the Public and Private HEI's in the Philippines

| E-readiness of Faculty and Students in Government and Private HEI's in the Philippines | Mean | | | Verbal Description |
|---|---|---|---|---|
| | Students | Faculty | Average | |
| State Universities and Colleges | 4.24 | 4.71 | 4.48 | Agree |
| Private Universities and Colleges | 3.96 | 4.09 | 4.03 | Agree |
| Overall Mean | 4.03 | 4.40 | 4.26 | Agree |

Table 2 shows the level of acceptability of faculty members and students from public and private universities in using e-learning in the educational process of teaching and learning. Students from both public and private universities showed a high level of acceptance in the use of e-learning based on the computed mean of 4.03 (public) and 4.06 (private) with a combined mean of 4.05, described as agree. In the same manner, faculty members showed in a high level of acceptance as supported by the computed mean of 4.44 (public) and 4.28 (private), with a combined mean of 4.36 described as agree. The overall mean is 4.21 described as agree. This finding indicates that both the faculty members and students from public and private universities believe that the use of e-learning can improve the quality of learning, increase their level of productivity, it is more effective than traditional tools and it is easy to use and learn.

Table 3 presents the perceived assessment of faculty members and students in terms of training. It is interesting to note that students from both public and private universities have a neutral stand in terms of training based on the computed mean of 2.69 (public) and 3.04 (private) with a combined mean of 2.87, described as neutral. The same sentiment was reflected in the assessment of the faculty, as faculty members from public and private universities registered a mean value of 2.63 and 2.71, respectively, and with a combined mean score of 2.67, described as neutral. The neutral response from both faculty and students was attributed to the insufficiency of pieces of training conducted in the universities in the use of e-learning infrastructure.

Table 2. Level of Acceptance of E-Learning among Faculty Members and Students in the Public and Private HEI's in the Philippines

| E-Learning Readiness of Faculty and Students in Government and Private HEI's in the Philippines | Mean | | | Verbal Description |
|---|---|---|---|---|
| | Students | Faculty | Average | |
| State Universities and Colleges | 4.03 | 4.44 | 4.24 | Agree |
| Private Universities and Colleges | 4.06 | 4.28 | 4.17 | Agree |
| Overall Mean | 4.05 | 4.36 | 4.21 | Agree |

Table 3. Training

| E-Learning Readiness of Faculty and Students in Government and Private HEI's in the Philippines in terms of Training | Mean | | | Verbal Description |
|---|---|---|---|---|
| | Students | Faculty | Average | |
| State Universities and Colleges | 2.69 | 2.63 | 2.66 | Neutral |
| Private Universities and Colleges | 3.04 | 2.71 | 2.88 | Neutral |
| Overall Mean | 2.87 | 2.67 | 2.77 | Neutral |

Table 4 presents the assessment of the faculty members and students from both public and private universities on the technological infrastructure. Students from both public and private universities assessed that the technological infrastructure has a positive impact on the adaptation of e-learning in the teaching-learning process based on the computed mean of 3.60 (public) and 3.47 (private), with a combined mean of 3.54 described as agreeing. This sentiment was also reflected on the part of the faculty members who gave a mean of 2.95 (public) and 3.48 (private), respectively, with a combined mean of 3.22 described as neutral.

The computed overall mean is 3.38 described as neutral. The result shows that the neutral response from the students and faculty was attributed to the inadequacy of technological infrastructure that will respond to the needs of faculty and students. Slow Internet connection, lack of laboratory facilities, and equipment and resources were among the major factors.

Table 5 shows the level of awareness of students and faculty members from public and private colleges in the Philippines. The result shows both the students from public and private universities are highly aware of using web technologies as a tool for e-learning. This was supported by the computed mean value of 4.72 for public and 4.65 for private universities with a combined mean of 4.69 described as highly aware. Similarly, faculty members from both public and private universities exhibited the same level of awareness based on the computed mean scores of 4.52 and 4.59 respectively.

The overall mean score is 4.62 described as highly aware. The high level of awareness from both the students and the faculty members was attributed to the fact that the students and teachers in the Philippines are utilizing web technology and resources in teaching and learning particularly in research and communication.

Table 4. Technological Infrastructure

| E-Learning Readiness of Faculty and Students in Government and Private HEI's in the Philippines in terms of Technological Infrastructures | Mean | | | Verbal Description |
| --- | --- | --- | --- | --- |
| | Students | Faculty | Average | |
| State Universities and Colleges | 3.60 | 2.95 | 3.28 | Neutral |
| Private Universities and Colleges | 3.47 | 3.48 | 3.48 | Neutral |
| Overall Mean | 3.54 | 3.22 | 3.38 | Neutral |

Table 5. Level of Awareness of Students and Faculty Members from Government and Private HEI's in the Philippines

| E-Learning Readiness of Faculty and Students in Government and Private HEI's in the Philippines in terms of Awareness | Mean | | | Verbal Description |
| --- | --- | --- | --- | --- |
| | Students | Faculty | Average | |
| State Universities and Colleges | 4.72 | 4.52 | 4.62 | Highly Aware |
| Private Universities and Colleges | 4.65 | 4.59 | 4.62 | Highly Aware |
| Overall Mean | 4.69 | 4.56 | 4.62 | Highly Aware |

There is no significant difference in the level of e-learning readiness based on the five factors among students from public and private colleges in the Philippines (Table 6). The computed t-values of 1.18 for e-readiness, 0.345 for acceptance, 1.195 for training, 0.429 for technological infrastructure, and 4-1.41 are all lower than the critical t-value of 2.02 at 0.05 level of significance. Thus, the null hypothesis was accepted. Therefore, the level of e-readiness of the students was the same regardless of whether they are attending public and private universities.

Table 7 shows that there is no significant difference in the level of e-learning readiness based on the four factors among faculty members from public and private colleges in the Philippines. The computed t-value of 1.26 for e-readiness, 1.94 for acceptance, 0.543 for training, and -6.61 for technological infrastructure, and 1.98 for awareness are all lower than the critical value of 2.02 at 0.05 level of significance. Therefore, the null hypothesis was accepted. This implies that faculty members from both public and private universities have the same level of e-learning readiness.

Table 6. Significant Difference in E-Learning Readiness Among Public and Private HEI Students

|  | Mean | | Computed t-value | Interpretation | Statistical Decision |
|---|---|---|---|---|---|
|  | Public | Private |  |  |  |
| E-Learning Readiness | 4.24 | 3.96 | 1.180 | NS | Accepted |
| Acceptance | 4.03 | 4.06 | 0.345 | NS | Accepted |
| Training | 2.69 | 3.04 | 1.195 | NS | Accepted |
| Technological Infrastructure | 3.60 | 3.47 | 0.429 | NS | Accepted |
| Awareness | 4.72 | 4.65 | -1.41 | NS | Accepted |
| df= 46; alpha = 0.05 | | | | | |

Table 7. Significant Difference in E-Learning Readiness Among Public and Private HEI'sFaculty Members

|  | Mean | | Computed t-value | Interpretation | Statistical Decision |
|---|---|---|---|---|---|
|  | Public | Private |  |  |  |
| E-Learning Readiness | 4.71 | 4.09 | 1.26 | NS | Accepted |
| Acceptance | 4.44 | 4.28 | 1.94 | NS | Accepted |
| Training | 2.63 | 2.71 | 0.543 | NS | Accepted |
| Technological Infrastructure | 2.95 | 3.48 | -6.61 | NS | Accepted |
| Awareness | 4.52 | 4.59 | 1.98 | NS | Accepted |
| df = 46; alpha= 0.05 | | | | | |

**CONCLUSION AND RECOMMENDATION**

Students from both public and private universities demonstrated a high level of e-learning readiness. Similarly, faculty from both public and private universities registered a high level of e-learning readiness. In terms of the acceptance on the use of e-learning, students and faculty members from public and private universities are all ready to adopt e-learning stressing that the use of e-learning can improve the quality of learning, increase their level of productivity, it is more effective than traditional tools and it is easy to use and learn. The neutral response from both faculty and students in terms of pieces of training was attributed to the insufficiency of pieces of training conducted in the universities in the use of e-learning infrastructure.

Students from both public and private universities assessed that the technological infrastructure has a positive impact on the adoption of e-learning in the teaching-learning process. In terms of awareness on the use of web technology as an e-learning tool, students and faculty members exhibited a high level of awareness. There is no significant difference in the level of e-learning readiness based on the four factors among students from public and private colleges in the Philippines. There is no significant difference in the level of e-learning readiness based on the four factors among faculty members from public and private colleges in the Philippines. thus, adoption and implementation of e-learning is recommended.